\documentclass[twoside]{photon2007}
\usepackage[latin1]{inputenc}
\usepackage[dvips]{graphicx,epsfig,color}
\usepackage{wrapfig,rotating}
\usepackage{amssymb,amsmath,array}

\begin{document}
\title{
Anomalous GPDs in the photon }
\author{S. Friot$^1$, B. Pire$^2$ and  L. Szymanowski$^{2, 3}$\thanks{L. Sz. is supported in part by the Polish Grant 1P03B02828}
\vspace{.3cm}\\
1-  Université Paris Sud-11 - Institut de Physique Nucléaire, \\
94510 Orsay, France  
\vspace{.1cm}\\
2-  Centre  de Physique Th{\'e}orique, \'Ecole Polytechnique, CNRS,
   \\
91128 Palaiseau, France 
\vspace{.1cm}\\
3-  Soltan Institute for Nuclear Studies,
Warsaw, Poland \\
}

\maketitle

\begin{abstract}
We consider deeply virtual Compton scattering (DVCS) on a photon target, 
in the generalized Bjorken limit, at the Born order and in the leading logarithmic approximation. This leads us to the extraction of the photon anomalous generalized parton distributions (GPDs) \cite{url, DVCSphoton}.
\end{abstract}

\section{Introduction}

At high energies, the point-like nature of the photon dominates over its hadronic component. This allows for a purely perturbative evaluation of its structure functions $F_2^\gamma$ and $F_L^\gamma$, which have therefore been the subject of much work during the seventies because, at this time, this unique feature was hoped to allow for an experimental confirmation of the fractional charge of quarks. Still at this time, the photon structure functions were also under intense study due to the completely different behaviour that they exhibit, compared to the nucleons structure functions. Indeed, it had been proven in \cite{Zerwas} that $F_2^\gamma$ shows a $\log Q^2$ behaviour already at the parton model (PM) level and, what was even more surprising, that this PM result is renormalized when taking into account QCD corrections \cite{Witten}.
A very accurate determination of the photon structure functions is now available \cite{Vogt} (see also \cite{Chyla}) and it is in accord with experiments \cite{Buras}.

On the other hand, for a bit more than ten years, one has been interested in the generalisation of DIS processes to exclusive hard reactions (see \cite{BL07} for the case with photons). This gave birth to new objects relevant for studies of the nucleons structure: the generalized parton distributions. DVCS on a real photon target looks like a theoretical laboratory for the study of the properties of GPDs. It may also become an interesting process for an electron-photon collider \cite{Krawczyk, Ginzburg}.

\section{\normalsize The DVCS process}

 Deeply virtual Compton scattering  on a photon target 
\begin{equation}\nonumber
\gamma^*(q) \gamma(p_1) \to \gamma(q') \gamma(p_2)
\label{dvcs}
\end{equation}
involves, at leading order in $\alpha_{em}$, and zeroth order in  $\alpha_{S}$, six one-loop Feynman diagrams: one box, one `crossed' and one `cat-ears'  with quarks in the loop (`quarks' diagrams), and three similar diagrams with antiquarks (`antiquarks' diagrams). 

The amplitude of the process $\gamma^*\gamma\rightarrow\gamma\gamma$ with one virtual and 
three real photons  can 
be written as
\begin{equation}\nonumber
A\doteq\, \epsilon_\mu\epsilon'^*_\nu{\epsilon_1}_\alpha{\epsilon^*_2}_\beta T^{\mu\nu\alpha\beta},
\end{equation}
where in our kinematics  the four
photon polarization vectors
$\epsilon(q)$, $\epsilon'^*(q')$, $\epsilon_1(p_1)$ and $\epsilon^*_2(p_2)$ are 
transverse.
The tensorial decomposition of $T^{\mu\nu\alpha\beta}$ reads\footnote{For simplicity, we restricted to the case where $\Delta_T$, the transverse part of
 $\Delta\doteq p_2-p_1$, vanishes, that is zero  scattering angle (but there is still a longitudinal momentum transfer).} \cite{BGMS75}
\begin{equation}\nonumber
T^{\mu\nu\alpha\beta} = \frac{1}{4}g^{\mu\nu}_Tg^{\alpha\beta}_T W_1+
\frac{1}{8}\left(g^{\mu\alpha}_Tg^{\nu\beta}_T 
+\right.g^{\nu\alpha}_Tg^{\mu\beta}_T
\end{equation}
\begin{equation}\nonumber
\left.
 -g^{\mu\nu}_Tg^{\alpha\beta}_T \right)W_2
+ \frac{1}{4}\left(g^{\mu\alpha}_Tg^{\nu\beta}_T - g^{\mu\beta}_Tg^{\alpha\nu}_T\right)W_3\, ,
\end{equation}
and it involves three scalar functions $W_i$, $i=1,2,3$.

The final result of our calculation of the DVCS amplitude can be expressed as an integral  over the quark momentum fraction $x$. It reads
\begin{equation}
 W_{1}  = \frac{e_q^4\,N_C}{2\,\pi^2}\int_{-1}^1\,dx
\frac{2\,x}{(x-\xi+i\eta)(x+\xi-i\eta)} \nonumber\\
\end{equation}
\begin{equation}
\times \left[\theta(x-\xi)  \frac {x^2 + (1-x)^2-\xi^2}{1-\xi^2}\; \right.\nonumber\\
 \end{equation}
 \begin{equation}
+\ \theta(\xi-x) \theta(\xi+x) \frac{x(1-\xi)}{\xi(1+\xi)} \nonumber\\
\end{equation}
\begin{equation}
\hspace{0.5cm}- \left.\theta(-x-\xi)  \frac {x^2 + (1+x)^2-\xi^2}{1-\xi^2}\;\right] \log\frac{m^2}{Q^2}
\label{W1}\ ,
\end{equation}
\begin{equation}
\hspace{-5.6cm} W_{2}  = 0
 \end{equation}
and
\begin{equation}
 W_{3}  = \frac{e_q^4\,N_C}{2\,\pi^2}\int_{-1}^1\,dx
\frac{2\xi}{(x-\xi+i\eta)(x+\xi-i\eta)}\nonumber\\
\end{equation}
\begin{equation}
 \left[\theta(x-\xi)   \frac {x^2 - (1-x)^2-\xi^2}{1-\xi^2}\; \right. \nonumber\\
\end{equation}
\begin{equation}
- \theta(\xi-x) \theta(\xi+x) \frac{1-\xi}{1+\xi} \;\nonumber\\
\end{equation}
\begin{equation}
\hspace{0.5cm}
+ \left. \theta(-x-\xi)  \frac {x^2 - (1+x)^2-\xi^2}{1-\xi^2}\;\right] \log\frac{m^2}{Q^2}\;.
\label{W3}
\end{equation}
At this point, there are two important things to know about the intermediate steps of the calculation. First of all, each Feynman diagram possesses an UV divergence. These divergences cancell when summing the `quarks' box, crossed and cat-ears diagrams contributions (obviously a similar cancellation also occurs when summing the `antiquarks' diagrams contributions). The second point concerns the cat-ears diagrams. Although it is crucial to include their contributions to cancell UV divergences, they do not lead to any $\log  \frac{m^2}{Q^2}$ terms, therefore the handbag dominance interpretation of the leading logarithmic 
result will be justified. 

We now want to interpret this result from the point of view of QCD factorization based on the operator product expansion, 
yet still in the zeroth order of the QCD coupling constant and in the leading logarithmic approximation.
 For this, we write for any function ${\cal F}(x,\xi)$ the obvious identity :
 \begin{equation}
 \hspace{-1.3cm}
{\cal F}(x,\xi) \; \log \frac{m^2}{Q^2} = {\cal F}(x,\xi) \;\log \frac{m^2}{M_{F}^2}\nonumber
\end{equation}
 \begin{equation}
\hspace{3.0cm} + {\cal F}(x,\xi) \;\log \frac{M_{F}^2}{Q^2}\;,
\label{stupid}
\end{equation}
where $M_{F}$ is an arbitrary factorization scale. We will show below that the first term with $\log \frac{m^2}{M_{F}^2}$
may be identified with the quark content of the photon, 
whereas the second term with $\log \frac{M_{F}^2}{Q^2}$
corresponds to the so-called photon content of the photon, coming from the  matrix element of
the two photon correlator $A_\mu(-\frac{z}{2}) A_\nu (\frac{z}{2})$ which contributes at the same order in
$\alpha_{em}$ as the quark correlator to the scattering amplitude. 
Choosing $M_{F}^2= Q^2$ will allow to express the DVCS amplitude only 
in terms of the quark content of the photon.

\section{\normalsize  QCD factorization of the DVCS amplitude on the photon}

To understand the results of Eqs.~(\ref{W1}-\ref{W3}) within the QCD factorization, we  first consider 
two quark non local correlators on the light cone and their matrix elements between real photon states :\newpage
\begin{equation}\nonumber
\hspace{-5cm}F^q = 
\end{equation}
\begin{equation}
\label{Fqa}
\int \frac{dz}{2\pi} e^{ixz}\langle \gamma(p')| \bar q(-\frac{z}{2}N)
 \gamma.N q(\frac{z}{2}N)|\gamma(p) \rangle 
\end{equation}
and
\begin{equation}\nonumber
\hspace{-5cm}\tilde F^q = 
\end{equation}
\begin{equation}\nonumber
\int \frac{dz}{2\pi} e^{ixz}\langle \gamma(p')| \bar q(-\frac{z}{2}N)
 \gamma.N \gamma^5 q(\frac{z}{2}N)|\gamma(p) \rangle \; ,
\end{equation}
where we note $N= n/n.p$ and where we neglected, for simplicity of notation, both 
the electromagnetic and the gluonic Wilson lines.

There also exists the photon 
 correlator $ F ^{N \mu}(-\frac{z}{2}N) F^{\nu N}(\frac{z}{2}N)$ (where $F^{N \mu} = N_\nu F^{\nu \mu}$), which mixes  with the quark operators \cite{Witten}, but 
contrarily to the quark correlator matrix element, the photonic one begins at order $\alpha_{em}^0$,
as seen  for instance in the symmetric case (where $Z\doteq \frac{z}{2}N$):
\begin{equation}
\nonumber
\int \frac{dz}{2\pi} e^{ixz}\langle \gamma(p_{2})| F ^{N \mu}(-Z) F^{\nu N}(Z) g_{T\mu\nu}|\gamma(p_{1}) \rangle
\end{equation}
\begin{equation}
= -  g_T^{\mu\nu}
\epsilon_\mu (p_{1})\epsilon^*_\nu(p_{2})(1-\xi^2) [\delta(1+x) + \delta(1-x)]\label{FF}.
\end{equation}
The quark correlator matrix elements, calculated in the lowest order of $\alpha_{em}$ and $\alpha_{S}$, 
suffer from ultraviolet divergences, which we regulate 
through the usual dimensional regularization procedure, with  $d= 4+2\epsilon$. 
We obtain (with  $\frac{1}{\hat\epsilon} = \frac{1}{\epsilon} +\gamma_{E}-\log 4\pi$)
\begin{equation}
\hspace{-5cm}F^q= \nonumber
\end{equation}
\begin{equation}
\frac{N_C\,e_{q}^2}{4\pi^2} g_T^{\mu\nu}\epsilon_\mu (p_{1})\epsilon^*_\nu(p_{2}) 
 \left[\frac{1}{\hat\epsilon} + \log{m^2}\right] F(x,\xi)\,,
\label{Fgam}
\end{equation}
with 
\begin{equation}\nonumber
 F(x,\xi) = \frac{x^2+(1-x)^2-\xi^2}{1-\xi^2}\theta(1>x>\xi) 
 \end{equation}
 \begin{equation}\nonumber
 -\frac{x^2+(1+x)^2-\xi^2}{1-\xi^2}\theta(-\xi>x>-1) 
 \end{equation}
 \begin{equation}\nonumber
 +\frac{x(1-\xi)}{\xi+\xi^2}\theta(\xi>x>-\xi)
\end{equation}
for the $\mu\leftrightarrow \nu$ symmetric (polarization averaged) part. 
A similar result for 
 the antisymmetric (polarized) part can be found in our original publication \cite{DVCSphoton}.

 The ultraviolet divergent parts
 are removed through the renormalization procedure  involving both quark and photon correlators. 
  The $\frac{1}{\hat \epsilon}$ terms in 
Eq.~(\ref{Fgam}) define the 
non-diagonal element $Z_{\bar q  q\;F
F}$ of the multiplicative matrix of  renormalization constants $Z$. 
The $\frac{1}{\hat \epsilon}$ terms are
then subtracted by the renormalization of quark operators.
The renormalization procedure introduces a renormalization scale which we here identify 
with a factorization scale  $M_{F }$ in the factorized form of the amplitude. Imposing the renormalization condition that the renormalized 
quark correlator matrix element vanishes
when the factorization scale $M_{F } = m$, we get from Eq.~(\ref{Fgam}) for the renormalized matrix element
(\ref{Fqa})
\begin{equation}\nonumber
F^q_{R} 
 = \frac{N_C\,e_{q}^2}{4\pi^2} g_T^{\mu\nu}\epsilon_\mu (p_{1})\epsilon_\nu(p_{2}) 
  \log{\frac{m^2}{M_{F}^2}} F(x,\xi) \; ,
\label{FRgam}
\end{equation}
and a similar result for the antisymmetric case.

These results permit us to define the generalized quark distributions in the photon, $H_i^q(x,\xi,0)$, as
\begin{equation}\nonumber
\hspace{-0.75cm}F^q_{R}= - g_T^{\mu\nu}\epsilon_\mu (p_{1})\epsilon^*_\nu(p_{2}) H_{1}^q (x,\xi,0)\;,
\end{equation}
\begin{equation}\label{Hquark}\nonumber
\tilde F^q_{R}=  i\epsilon^{\mu\nu p N}\epsilon_\mu (p_{1})\epsilon^*_\nu(p_{2}) H_{3}^q (x,\xi,0)\;
\end{equation}
and to 
write the quark contribution to the DVCS amplitude as a convolution of coefficient functions and 
distributions $ H_{i}^q$  
\begin{equation}\nonumber
 \hspace{-0.6cm}W^q_{1}= \int\limits_{-1}^1 dx\ C_V^q(x)  H_{1}^q (x,\xi,0)\;\,,
 \end{equation}
\begin{equation}
W^q_{3}= \int\limits_{-1}^1 dx\  
C_A^q(x)  H_{3}^q (x,\xi,0)\; ,
\label{fac}
\end{equation}
where the  Born order coefficient functions $C_{V/A}^q$ attached to the quark-antiquark symmetric and 
antisymmetric correlators are the usual hard process amplitudes :
 \begin{equation}\nonumber
 C_{V/A}^q = - 2e_q^2\left(\frac{1}{x-\xi+i\eta} \pm \frac{1}{x+\xi-i\eta}\right)\; .
\end{equation}
We recover in that way the $\log \frac{m^2}{M_{F}^2}$ term in the right hand side of Eq.~(\ref{stupid}).

The photon operator contribution to the DVCS amplitude at the order $\alpha_{em}^2$ considered here,
involves  a new coefficient function calculated at the factorization scale $M_{F}$, which
 plays the role of the infrared cutoff, convoluted with the
photon correlator (\ref{FF}) or with its antisymmetric counterpart. 
The results of these convolutions effectively coincide
with the amplitudes calculated in Section 2 with the quark mass replaced by the factorization scale 
$m \to M_{F}$, and leads to the second term in the right hand side of Eq.~(\ref{stupid}).
The triviality of Eq.~(\ref{stupid}) in fact hides the more general independence of the scattering amplitude
 on the choice of the scale  $M_{F}$ which is controlled by the renormalization group equation.

We still have the freedom to fix the factorization scale $ M_{F}^2$ in any convenient way.
Choosing $ M_{F}^2 = Q^2$ kills the logarithmic terms coming from the photon correlator, so that the 
DVCS amplitude is written (at least, in the leading logarithmic approximation) solely in terms of the quark correlator,
recovering a partonic interpretation of the process.

\section{Conclusion}

We derived the leading amplitude of the DVCS process on
a photon target. We have shown that the amplitude coefficients $W_i^q$ factorize in the forms shown in 
Eq.~(\ref{fac}),
irrespectively of the fact that the handbag diagram interpretation appears only 
{\em after} cancellation of UV divergencies in the scattering amplitude.
We have shown that the objects $H_i^q(x,\xi,t)$ are matrix elements of non-local quark operators 
on the light cone, and that they have an anomalous component which is proportional to $\log(Q^2/m^2)$.
They thus have all the properties attached to generalized parton distributions.

This work is the first step towards the study of the anomalous generalized parton distributions  in the photon. It would be nice to consider the $\Delta_T\neq0$ kinematics. Another logical step beyond this analysis is the inclusion of QCD corrections. Since we have demonstrated the presence of anomalous terms both in the DGLAP and ERBL regions, this can be done by solving the non-homogeneous DGLAP-ERBL evolution equations, which are the generalisation of the non-homogeneous DGLAP equations for the photon structure functions.

\end{document}